\documentclass[aps,onecolumn,11pt,floatfix,altaffilletter,tightenlines,showpacs,showkeys,notitlepage,nofootinbib,preprintnumbers]{revtex4-1}
\pdfoutput=1
\usepackage[dvipdf,dvips,pdftex]{graphicx}
\usepackage[colorlinks=true,citecolor=blue,linkcolor=blue]{hyperref}
\usepackage{amsmath}
\usepackage{amssymb}
\usepackage{epsfig}
\usepackage{epstopdf}
\usepackage{url}
\usepackage{color}
\usepackage[utf8]{inputenc}
\usepackage{empheq}
\usepackage{here}
\usepackage{bm}
\usepackage{braket}
\usepackage{comment}

\newcommand{\be}{\begin{eqnarray}}
\newcommand{\ee}{\end{eqnarray}}

\catcode`\@=11

\def\lsim{\mathrel{\mathpalette\@versim<}}
\def\gsim{\mathrel{\mathpalette\@versim>}}
\def\@versim#1#2{\vcenter{\offinterlineskip
\ialign{$\m@th#1\hfil##\hfil$\crcr#2\crcr\sim\crcr } }}
\catcode`\@=12

\begin{document}

\title{Electroweak baryogenesis between broken phases in multi-step phase transition}

\author{Mayumi Aoki$^{1\,}$} \email{mayumi.aoki@staff.kanazawa-u.ac.jp}
\author{Hiroto Shibuya$^{1\,}$} \email{h$\_$shibuya@hep.s.kanazawa-u.ac.jp}

\affiliation{%
$^1$Institute for Theoretical Physics, Kanazawa University,
Kanazawa 920-1192, Japan
}

\pagestyle{plain}

\preprint{KANAZAWA-23-03}

%
\begin{abstract}
Possibility of electroweak baryogenesis (EWBG) via multi-step phase transition (PT) is considered. We investigate the EWBG between $SU(2)$ broken phases in the second step PT of the two-step PT. The produced baryon number asymmetry is evaluated using a prototypical model with a $SU(2)$ charged CP-odd scalar field. We show that the second step PTs where the sphaleron rate is possible to be un-suppressed before the PTs, which can reproduce the sufficient baryon number asymmetry. Our studies can be adapted to models with extra $SU(2)$ scalar fields. We discuss specific models suitable for our scenario. 
\end{abstract}
\maketitle

%

\vspace{0.2in}
\section{Introduction}\label{sec:Introduction}
The baryon asymmetry in the universe (BAU) is one of the big problems in particle physics.
The value of the BAU $\eta_B\equiv(n_B-n_{\bar B})/s$ is observed by the Planck satellite as~\cite{Aghanim:2018eyx}
\begin{equation}\label{eq:baobs}
\eta_B^{\rm obs}=(8.72\pm 0.08)\times10^{-11} \,,
\end{equation}
where $n_B~(n_{\bar B})$ is the number density of (anti-)baryon number $n_B~(n_{\bar B})$ and $s$ is the entropy density. This value is consistent with the observation of the Big Bang nucleosynthesis~\cite{ParticleDataGroup:2020ssz}.
The standard model (SM) cannot produce the asymmetry because it does not satisfy the conditions for achieving the BAU. The conditions are called Sakharov conditions~\cite{Sakharov:1967dj}, and the contents are (i) baryon number violation, (ii) C and CP violation, and (iii) departure from equilibrium.
Those conditions need to be satisfied when producing the BAU.
The SM fulfills the first condition but does not sufficiently do the second and third ones.
The SM has the CP phase called the Cabibbo-Kobayashi-Maskawa phase in the Yukawa matrix.
It is known that the phase is too small to explain the observed BAU~\cite{Gavela:1993ts, Huet:1994jb, Gavela:1994dt}.
The third condition, a departure from equilibrium, can be satisfied via the first-order electroweak phase transition (EWPT) in the SM if the Higgs boson mass $m_h$ is lower than about 70 GeV~\cite{Kajantie:1995kf,Csikor:1998eu}.
Nevertheless, because of $m_h=125$ GeV, the EWPT becomes the cross-over, and the discontinuous PT cannot occur~\cite{DOnofrio:2014rug}.
From above, the SM cannot explain the BAU.

One of the extensions to realize the observed BAU is expanding the scalar sector of the SM.
There are various ways to expand by inserting a scalar singlet (the singlet extension), a doublet (two Higgs doublet models (2HDMs)), and a triplet (the triplet extension) into the SM.
We can combine these models and consider their extensions.
In the above multi-Higgs models, the first-order PT can occur contrary to the SM.
Adequate CP-violating (CPV) sources can also be achieved by considering high-dimensional operators or CPV sources in the scalar potential.
Hence, the EWBG can be realized in the extended models.
On the other hand, the multi-step PTs can occur in the multi-Higgs models because the extra scalar fields which can get vacuum expectation values (VEVs) exist.
The features of the multi-step PTs are investigated in Refs.~\cite{Patel:2012pi, Blinov:2015sna, Inoue:2015pza, Hashino:2018zsi, Chiang:2019oms, Aoki:2021oez, Benincasa:2022elt}. They can change the calculation of the dark matter (DM) production as shown in Refs.~\cite{Bian:2018mkl, Bian:2018bxr, Elor:2021swj, Azatov:2022tii, Shibuya:2022xkj}.
In addition, the multi-step PTs can achieve EWBG and Refs.~\cite{Jiang:2015cwa, Cline:2021iff, Huber:2022ndk} calculated the produced BAU via the multi-step in the singlet extension. 
However, in models with extra $SU(2)_L$ scalar fields, the produced BAU via the multi-step PT has not been calculated specifically.

In this letter, we investigate the EWBG in a prototypical model with an extra $SU(2)_L$ scalar field and calculate the produced baryon number asymmetry.
The applicable models are the 2HDMs, the triplet scalar extension as shown in Ref.~\cite{Chala:2018opy}, and those extensions.
We consider that the second step in the two-step PT produces the BAU to change the CP phase during the PT.
In contrast, the EWBG via the first step in the multi-step PT is proposed in Ref.~\cite{Patel:2012pi} and discussed in, e.g., Refs.~\cite{Inoue:2015pza, Blinov:2015sna, Aoki:2021oez}.
However, the CP phase is difficult to change in the first step PT if there are no minima or saddle points which make the path of the PT curved.
Hence, we do not consider such cases.
Since the second step PT occurs from the $SU(2)_L$ broken phase to another broken phase, the EW sphaleron rate may be suppressed. 
In this letter, we reveal that even if the outside of the bubble is the broken phase, the sphaleron rate is hardly suppressed, and adequate baryon asymmetry can be accomplished in certain cases.
The above scenario was suggested in Ref.~\cite{Land:1992sm} considering the simplified 2HDM. Its difficulty due to the sphaleron suppression was represented in Ref.~\cite{Hammerschmitt:1994fn}.
Although Ref.~\cite{Hammerschmitt:1994fn} used the simplified model and PT calculation, we investigate two-step PT in a more general model called the inert doublet model, including loop contributions for the scalar potential.
As a result, we find the two-step PTs which would have enough sphaleron rate before the second step PT occurs in contrast to Ref.~\cite{Hammerschmitt:1994fn}.

This letter consists of the following sections. Section~\ref{sec:2} is devoted to describing the multi-step PT we consider.
In Section~\ref{sec:3}, we discuss how to include the sphaleron rate in the $SU(2)_L$ broken phase and show the equations for evaluating the produced BAU.
We give numerical results and discuss the possibility of achieving adequate BAU in Section~\ref{sec:4}.
We also discuss specific models suitable for our scenario and give the conclusion in Section~\ref{sec:5}.

\section{Multi-step electroweak phase transition}\label{sec:2}
\begin{figure}[t]
\begin{center}
 \includegraphics[width=6cm]{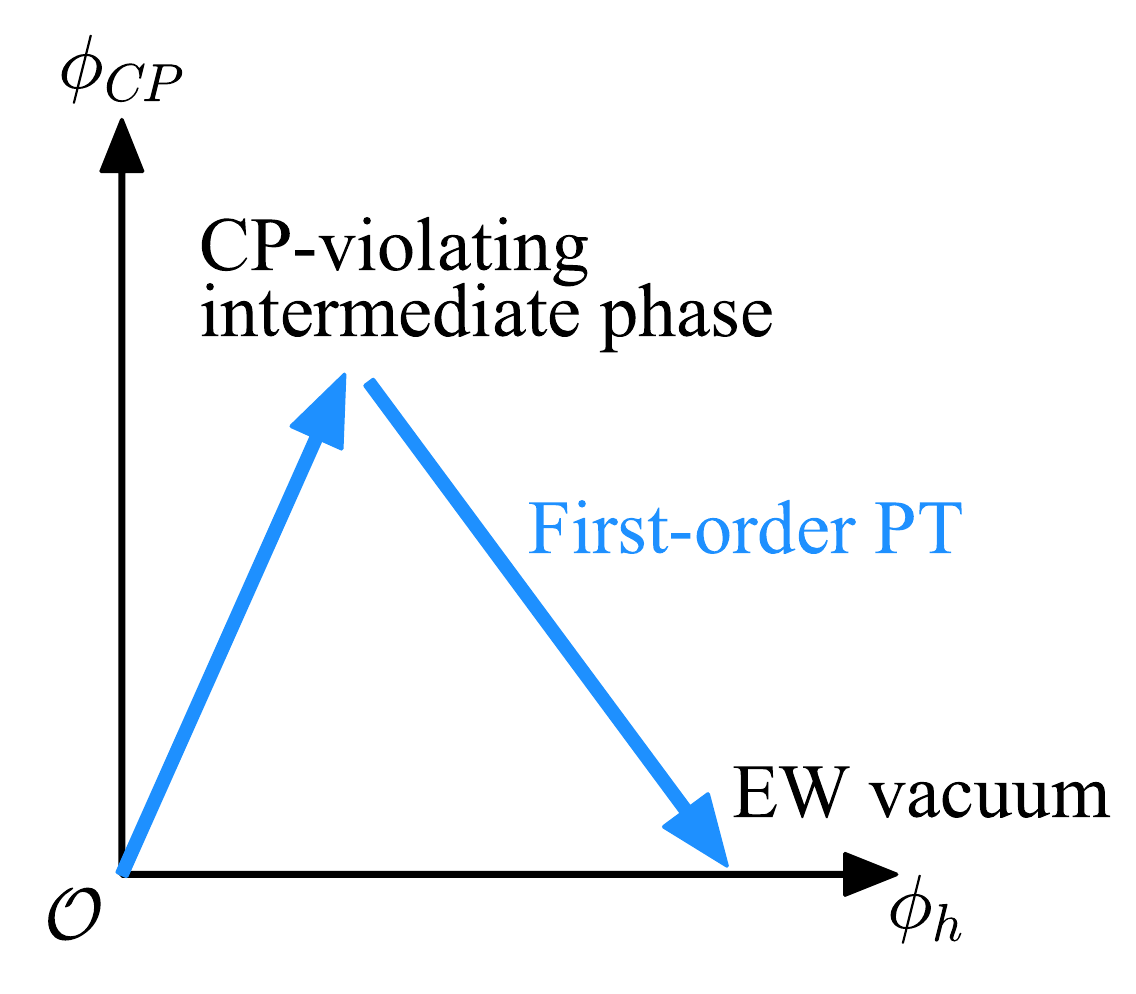}
\caption{Schematic picture of the multi-step transition path in this letter. In the first step, the PT occurs from the origin to the CPV intermediate phase. After the temperature decreases, the second step happens from the intermediate phase to near the EW vacuum. Only the second step needs to be first-order and produces the BAU.}
\label{fig:schematicBG}
\end{center} 
\end{figure}

To demonstrate the produced baryon asymmetry via the multi-step PTs, we consider the following prototypical model with $SU(2)_L$ charged scalar fields.
The complex top quark mass, which induces the CP violation, is written as
\begin{align}\label{eq:complexmt}
m_t = \frac{1}{\sqrt{2}}y_t(\phi_h+i\phi_{_{CP}})\equiv
\frac{1}{\sqrt{2}}y_t\phi e^{i\theta_{_{CP}}},
\end{align}
with the top Yukawa coupling $y_t$, the neutral CP-even and CP-odd scalar fields that have $SU(2)_L$ charges, $\phi_h$ and $\phi_{_{CP}}$, respectively, and $\phi\equiv \sqrt{\phi_h^2+\phi_{_{CP}}^2}$.
The CP phase $\theta_{_{CP}}$ is simply determined by the complex phase of $\phi_h+i\phi_{_{CP}}$.
We consider a multi-step EWPT shown in Fig.~\ref{fig:schematicBG}.
In this scenario, 
the first step PT occurs from the origin to a minimum with $\phi_{_{CP}}\neq0$.
This PT can be continuous or non-continuous.
The CP phase does not change in this PT because of the straight path from the origin, and no BAU is produced.
After the temperature decreases, 
the second step PT happens from the minimum with $\phi_{_{CP}}\neq0$ to near the EW vacuum, $(\phi_h, \phi_{_{CP}})=(v, 0)$ with $v= 246$ GeV.
The second step PT needs to be first-order to make the BAU.
If the first step PT occurs along the $\phi_{_{CP}}$ axis with $\phi_h=0$, the CP phase changes dramatically during the second PT, $\theta_{_{CP}}=\pi/2\rightarrow0$. This makes the amount of the produced BAU large.

We assume that the scalar VEVs before the second step PT as $(\phi_h, \phi_{_{CP}})=(w_h, w_{_{CP}})$ and bubble profiles in the second step PT can be written by kink solutions as
\begin{equation}
\phi_h(z)=\frac{(v-w_h)}{2}\left(1-\mbox{tanh}\frac{z}{L_w}\right)+w_h\,,\quad
\phi_{_{CP}}(z)=\frac{w_{_{CP}}}{2}\left(1+\mbox{tanh}\frac{z}{L_w}\right)\,,
\label{eq:phikink}
\end{equation}
where $z$ is the radial coordinate of the bubble in the wall frame, and the bubble wall locates at $z=0$. The index $L_w$ shows the bubble wall width.

\section{Sphaleron rates and baryon asymmetry}\label{sec:3}
The CP phase in the fermion mass raises CP-violating interactions between the fermions and the bubble wall.
The interactions create the non-zero chemical potential for the left-handed baryons $\mu_{B_L}$.
The chemical potential changes the baryon number via the EW sphaleron.
Assuming the sphaleron processes are suppressed only inside the bubble,
the chemical potential outside of the bubble changes to the net baryon asymmetry via the $SU(2)$ sphaleron process. 

The sphaleron rate in the $SU(2)$ symmetric phase is given by the dimensional grounds as~\cite{Arnold:1987mh, Khlebnikov:1988sr}
\begin{align}
&\Gamma_{\rm sph}^{\rm sym}=k\alpha_{_W}^4T\sim 10^{-6}T\,,
\label{eq:gamma_sym}
\end{align}
where $k$ is evaluated numerically as $0.1\lesssim k\lesssim1$~\cite{Ambjorn:1995xm, Quiros:1999jp} and $\alpha_{_W}=g^2/(4\pi)\simeq0.034$.
The coupling $g$ represents the $SU(2)$ gauge coupling.
On the other hand, the sphaleron rate in the $SU(2)$ broken phase contains the exponential suppression factor and has been calculated as~\cite{Arnold:1987mh}
\begin{align}
&\Gamma_{\rm sph}^{\rm br}\sim 2.8\times 10^5 T \left(\frac{\alpha_{_W}}{4\pi}\right)^4\kappa\left(\frac{\xi}{B}\right)^7e^{-\xi}\,,
\label{eq:gamma_broken}
\end{align}
with the index $\xi(T)=E_{\rm sph}(T)/T$ and the functional determinant $\kappa$ estimated as $10^{-4}\lesssim \kappa\lesssim10^{-1}$~\cite{Dine:1991ck}.
The sphaleron energy $E_{\rm sph}$ is given by~\cite{Quiros:1999jp}
\begin{align}
&E_{\rm sph}(T)=\frac{g\phi(T)}{\alpha_{_W}}B\,,
\end{align}
where the constant $B$ is evaluated as the function of $x=m_h/m_{_W}$,
\begin{align}
&B(x)=1.58 + 0.32x - 0.05x^2\,.
\end{align}
In the SM, the value of $B$ is $B\simeq1.96$.

\begin{figure}[t]
\begin{center}
\includegraphics[width=7.5cm]{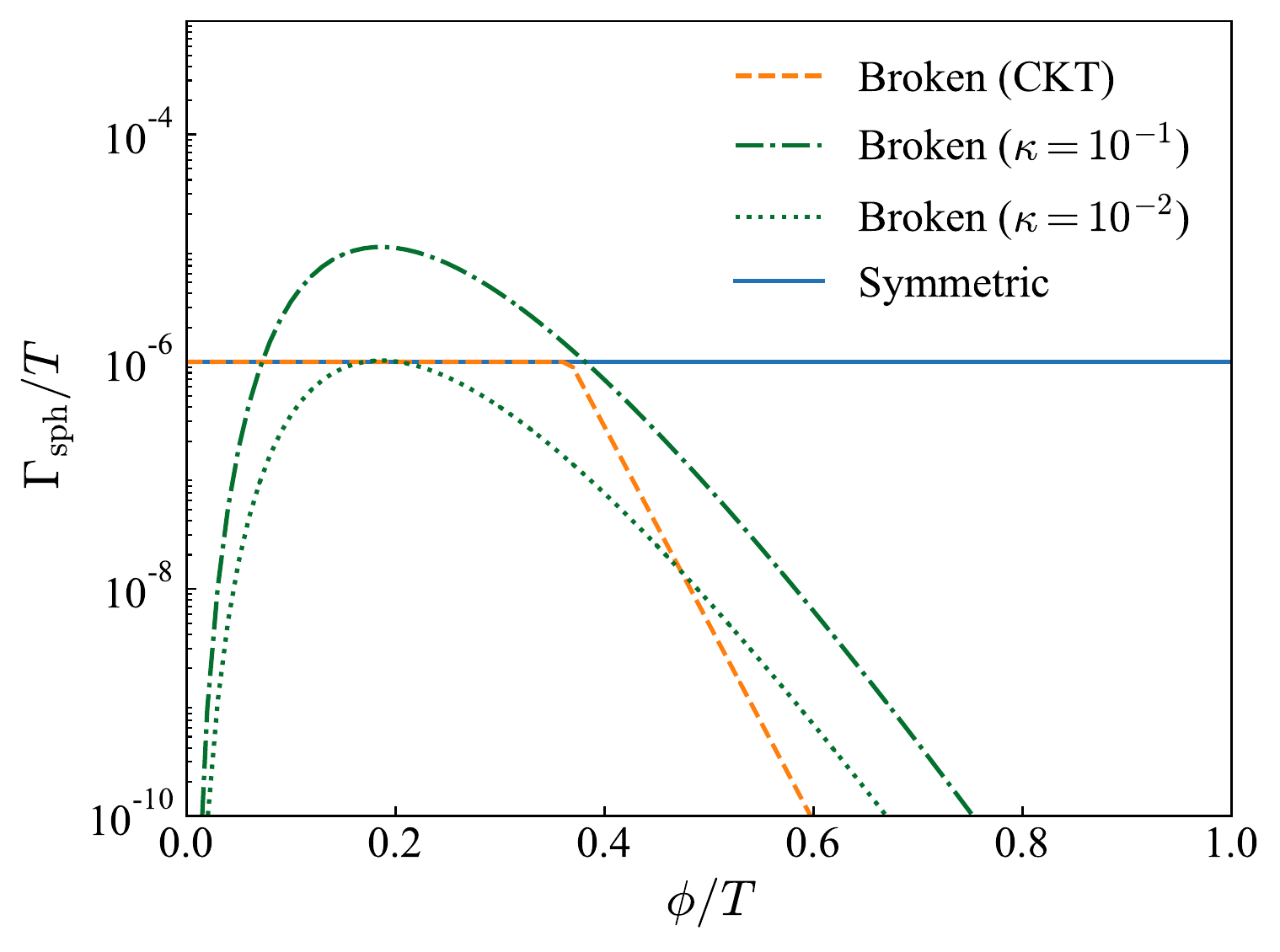}
\caption{
Sphaleron rates $\Gamma_{\rm sph}/T$ in the CKT prescription $\Gamma_\mathrm{\rm sph}^{\rm sym} f_\mathrm{sph}/T$ (orange), the broken phase $\Gamma_{\rm sph}^{\rm br}/T$ (green), and the symmetric phase $\Gamma_{\rm sph}^{\rm sym}/T$ (blue) in the function of $\phi /T$. We take $\kappa=10^{-1}$ (green dash-dot) and $10^{-2}$ (green dotted) in the broken phase case.
}
\label{fig:GammaSphleron}
\end{center}
\end{figure}

In order to treat a region where the calculation of $\Gamma_{\rm sph}^{\rm br}$ is not reliable, Ref.~\cite{Cline:2011mm} suggested treating the overall sphaleron by introducing the suppression factor $f_\mathrm{sph}$ as $\Gamma_\mathrm{\rm sph}^{\rm sym} f_\mathrm{sph}$, where the factor is given by
\begin{equation}
f_\mathrm{sph}(z) = \mathrm{min}
\left[ 1, \ \frac{ 2.4 T }{ \Gamma_\mathrm{\rm sph}^{\rm sym} } \exp\left(\frac{-40 \phi(z) }{T}\right) \right]\,.
\end{equation}
In this letter, we call the method the CKT prescription.
In Fig.~\ref{fig:GammaSphleron}, we show the sphaleron rates $\Gamma_{\rm sph}/T$ in the CKT prescription (orange), the broken phase (green), and the symmetric phase (blue) in the function of $\phi /T$. We take $\kappa=10^{-1}$ (green dash-dot) and $10^{-2}$ (green dotted) in the broken phase case.
We can see that the rate in the CKT prescription is the same as that in the symmetric phase for $\phi /T\lesssim0.4$. Afterward, the rate in the CKT rapidly decreases by the exponential suppression.
For very small $\phi/T$, $\Gamma_{\rm sph}^{\rm br}$ approaches zero and not match with $\Gamma_{\rm sph}^{\rm sym}$. 
It is because the condition imposed when deriving the rate in Eq.~(\ref{eq:gamma_broken}), $\alpha_3\equiv gT/(4\pi \phi )\ll 1$ where $\alpha_3$ is the coupling constant of the three-dimensional theory, is severely broken in the region. We do not consider such a region.
We can also see the sphaleron rates with $\kappa=10^{-1}$ exceeds the rates in the symmetric phase in the region $0.1\lesssim\phi /T\lesssim0.4$ in Fig.~\ref{fig:GammaSphleron}.
Ref.~\cite{Arnold:1987mh} insists that the excess would be the consequence of the breakdown of the approximation.
The specific values of $\alpha_3$ are about $0.13$ for $\phi /T=0.4$ and $0.26$ for $\phi /T=0.2$.
To obtain a more precise sphaleron rate in the broken phase, non-perturbative calculations are needed.

The asymmetry between baryons and anti-baryons via the CPV interactions can be calculated by the transport equations.
Considering the top transport with the WKB method, there are two ways to calculate, the FH scheme \cite{Fromme:2006wx, Fromme:2006cm} and CK scheme \cite{Cline:2020jre}.
We use the latter scheme since it does not use the small bubble wall velocity approximation and is the more precise method.
The transport equations and the chemical potential for the left-handed baryons $\mu_{B_L}$ are described in Appendix~\ref{sec:A}.
We calculate the baryon asymmetry $\eta_B$ with the CKT prescription as
\begin{align}
\eta_B&=\frac{405\Gamma_{\rm sph}^{\rm sym}}{4\pi^2\gamma_wv_wg_*T}\int_0^{\infty} dz ~\mu_{B_L}f_\mathrm{sph} {\rm exp}\left(-\frac{45\Gamma_{\rm sph}^{\rm sym}f_\mathrm{sph} z}{4\gamma_wv_w}\right) 
\,, \label{eq:etaCK}
\end{align}
with the effective degrees of freedom of the plasma given by $g_*=106.75$ in the SM, the bubble wall velocity $v_w$, and $\gamma_w=1/\sqrt{1-v_w^2}$.
Refs.~\cite{Cline:2011mm, Cline:2020jre} insert the factor $f_\mathrm{sph}$ only to $\Gamma_{\rm sph}^{\rm sym}$ outside of the exponential. 
In our calculation, we include the factor to $\Gamma_{\rm sph}^{\rm sym}$ also in the exponential, although results do not change largely after including. 
We will also derive the baryon asymmetry $\eta_{B}^{{\rm br}}$ using the rate $\Gamma_{\rm sph}^{\rm br}$ in Eq.~(\ref{eq:gamma_broken}) as
\begin{align}
\eta_{B}^{{\rm br}}&=\frac{405\Gamma_{\rm sph}^{\rm br}}{4\pi^2\gamma_wv_wg_*T}\int_0^{\infty} dz ~\mu_{B_L}(z)\exp\left(-\frac{45\Gamma_{\rm sph}^{\rm br}z}{4\gamma_wv_w}
\right) \,, \label{eq:etaCKbr}
\end{align}
with $\kappa=10^{-1}$ and $10^{-2}$.
In the next section, we show results using the evaluation with the CKT prescription and $\Gamma_{\rm sph}^{\rm br}$ in Eqs.~(\ref{eq:etaCK}) and (\ref{eq:etaCKbr}), respectively.

\section{Numerical results}\label{sec:4}
\begin{figure}[t]
 \begin{center}
 \includegraphics[width=7.5cm]{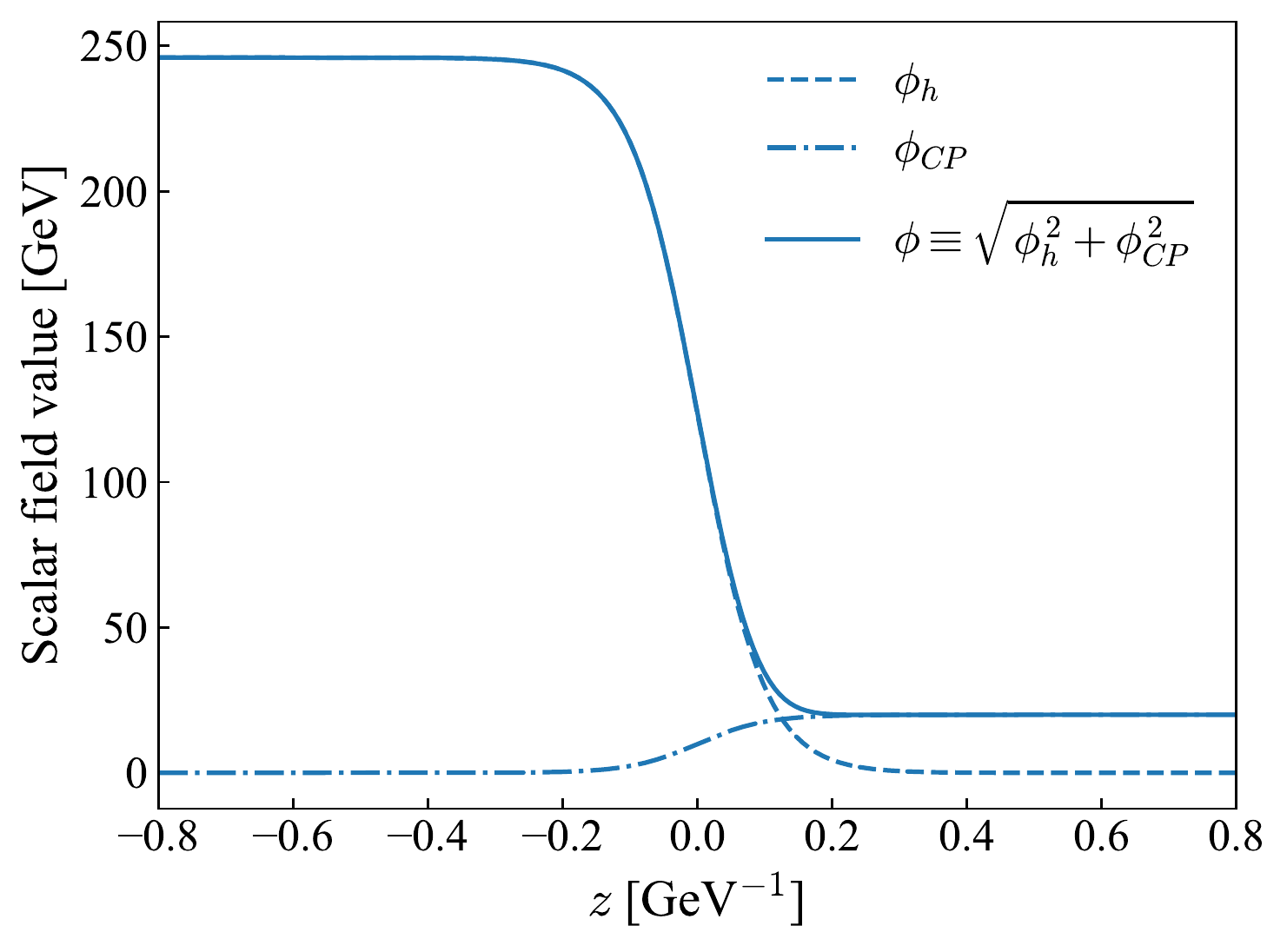}
 \includegraphics[width=7.5cm]{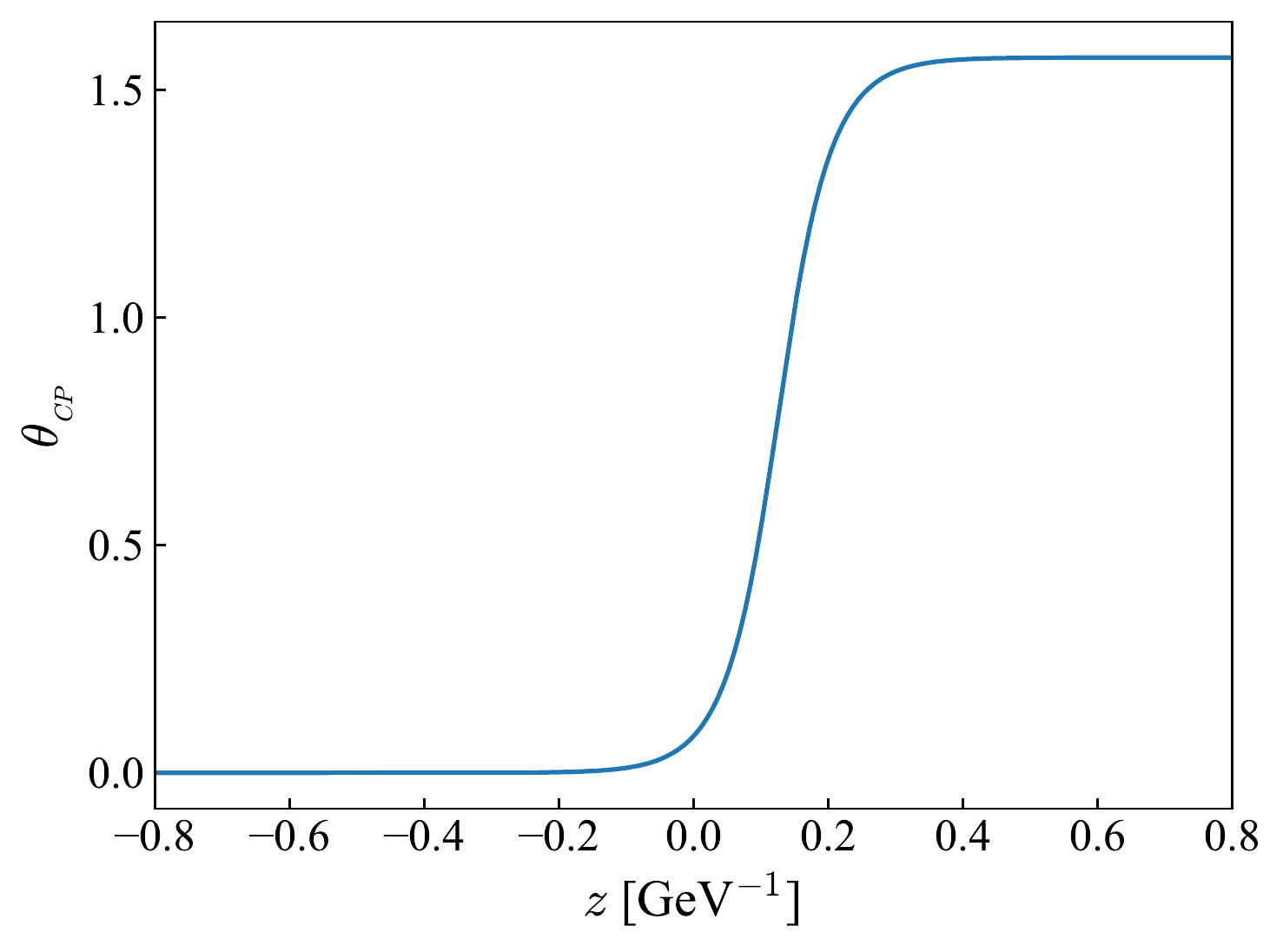}
\caption{Bubble profiles for the scalar fields (left) and the CP phase (right) of the second step PT in the benchmark scenario.}
\label{fig:bubbleprofile}
 \end{center}
\end{figure}
\begin{figure}[t]
 \begin{center}
 \includegraphics[width=7.5cm]{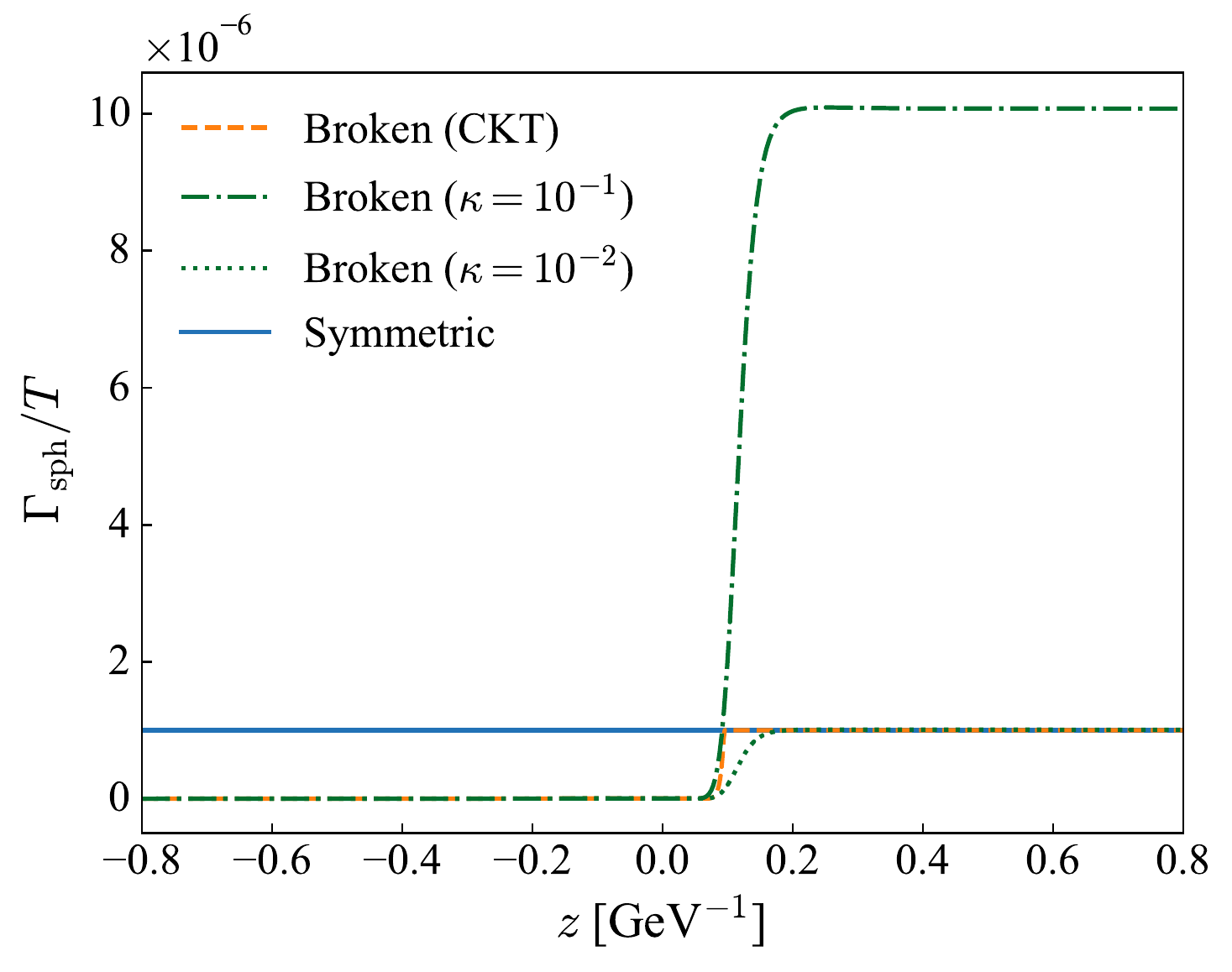}
 \includegraphics[width=7.5cm]{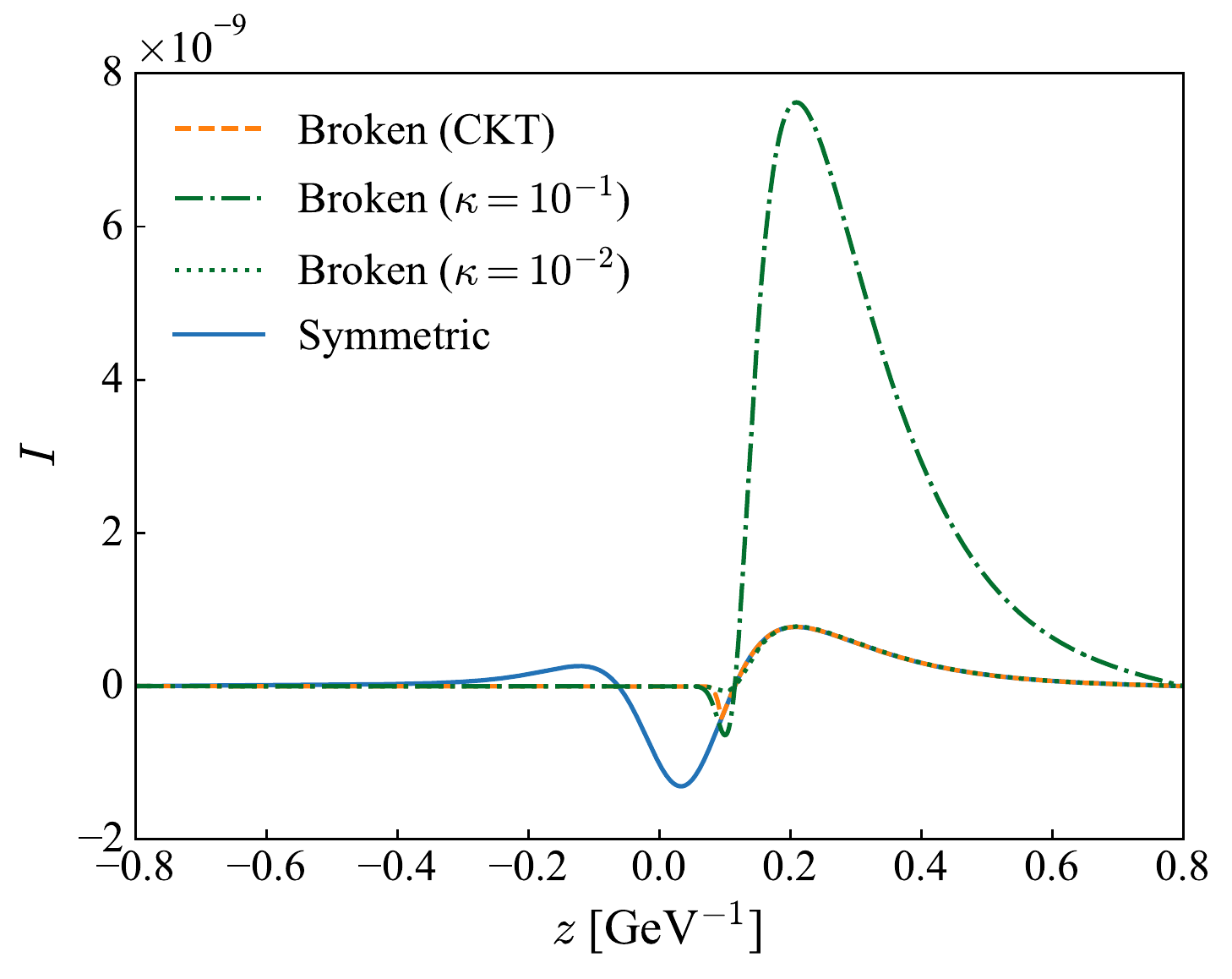}
\caption{
Sphaleron rates $\Gamma_{\rm sph}/T$ (left) and integrands $I$ (right) determined by $\eta_B=\int_0^\infty dz~I(z)$ with the rate in the CKT prescription (orange), the broken phase (green), and the symmetric phase (blue) as the function of $z$ in the benchmark scenario.
We take $\kappa=10^{-1}$ (green dash-dot) and $10^{-2}$ (green dotted) in the broken phase case.
}
\label{fig:integrand}
 \end{center}
\end{figure}
\begin{figure}[t]
 \begin{center}
 \includegraphics[width=7.5cm]{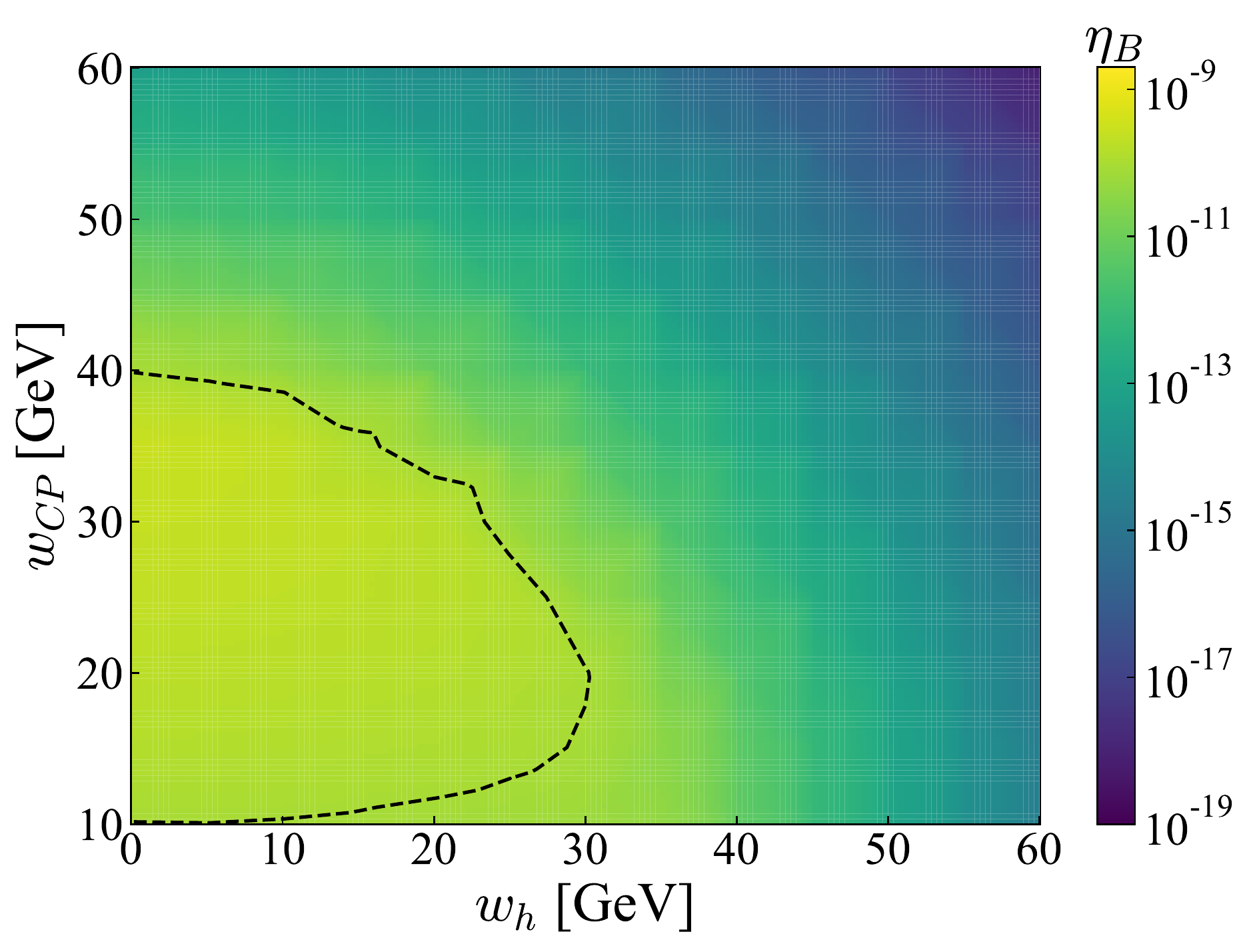}
\caption{Produced baryon number asymmetry in the $w_h$ vs. $w_{_{CP}}$ plane by using the CKT prescription.
The colors indicate the produced asymmetries when the second step PT occurs from the point $(w_h, ~w_{_{CP}})$ to the EW vacuum.
The black dashed line shows the observed value, $\eta_B^{\rm obs}\simeq8.72\times 10^{-11}$.
}
\label{fig:baryonasymmetery}
 \end{center}
\end{figure}
\begin{figure}[t]
 \begin{center}
 \includegraphics[width=7.5cm]{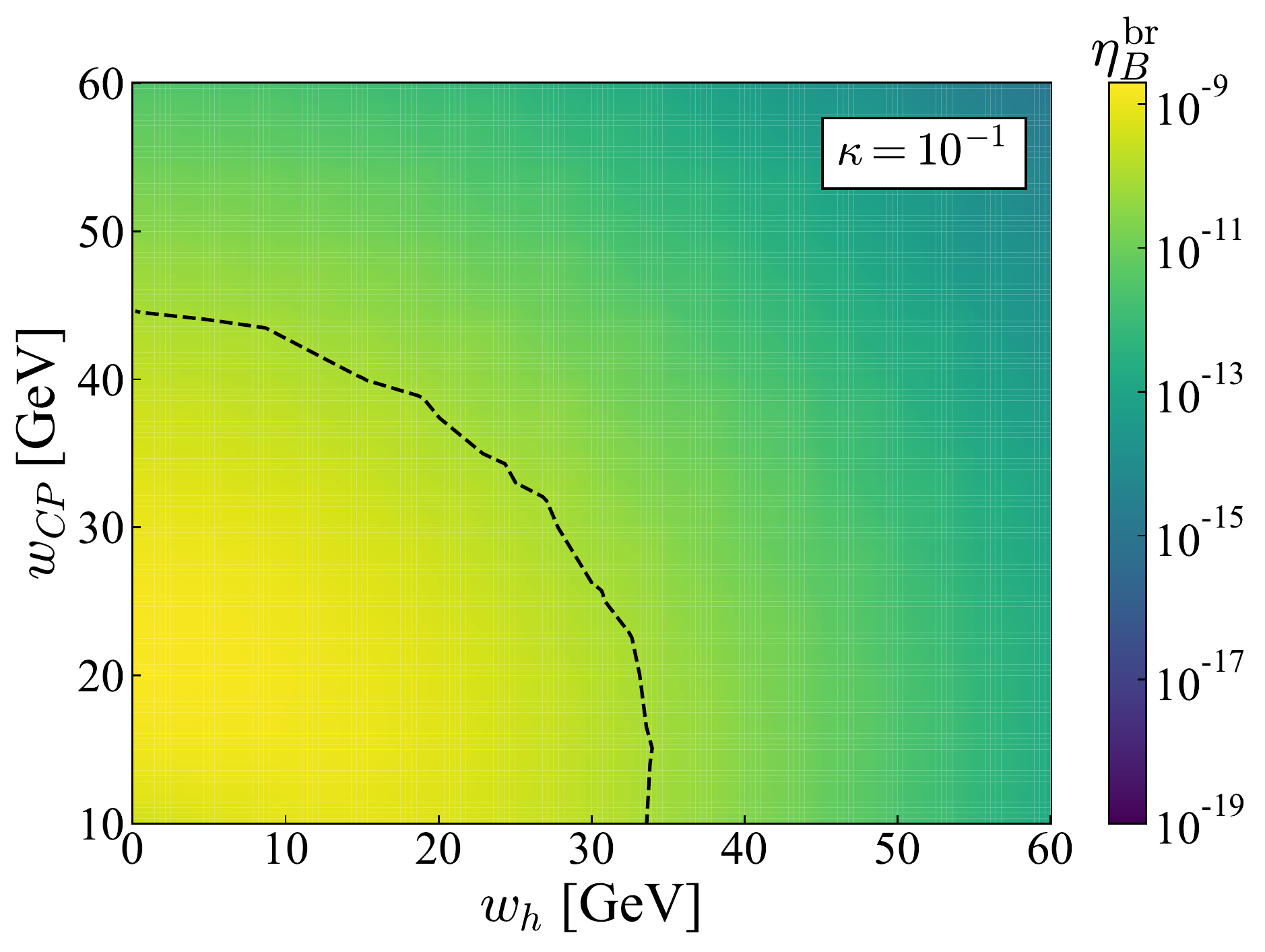}
 \includegraphics[width=7.5cm]{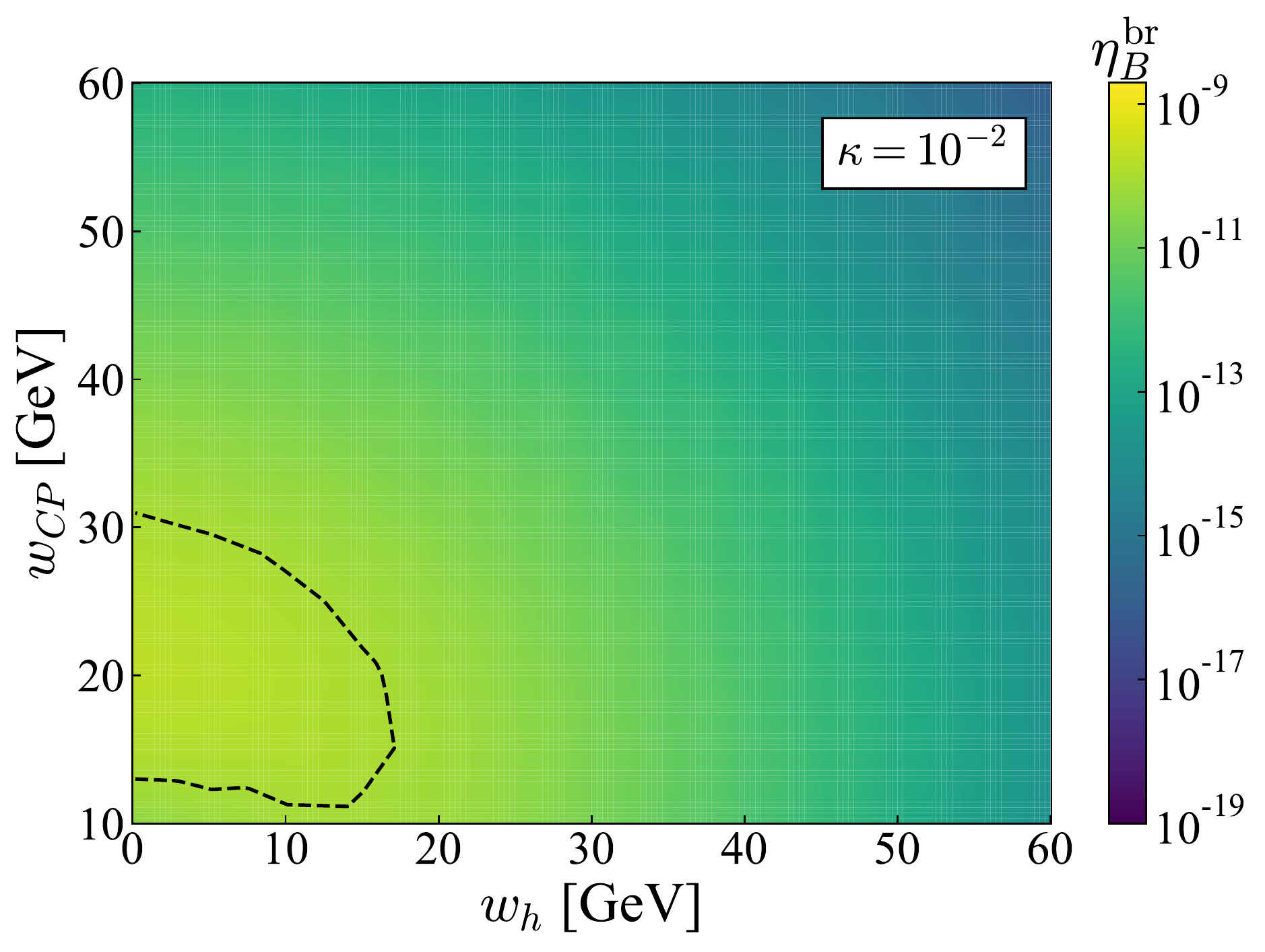}
\caption{
Same plots with Fig.~\ref{fig:baryonasymmetery} by using $\Gamma_{\rm sph}^{\rm br}$ with $\kappa=10^{-1}$ (left) and $\kappa=10^{-2}$ (right).}
\label{fig:baryonasymmetery2}
 \end{center}
\end{figure}

In this section, we show produced baryon number asymmetries based on the calculation we introduced.
To demonstrate the asymmetries, we set a benchmark scenario for the second step PT as
\begin{align}
&(\phi_h, ~\phi_{_{CP}})=(0,~20~{\rm GeV})~\rightarrow~(246~{\rm GeV},~0)\,,\label{eq:BMPT}
\end{align}
with the bubble nucleate temperature $T_n=100$ GeV.
The values of bubble wall velocity $v_w$ and the bubble width $L_w$ are chosen as 0.1 and 0.1 GeV$^{-1}$, respectively. 
In that case, the bubble profiles for the scalar fields in Eq.~(\ref{eq:phikink}) and the CP phase $\theta_{_{CP}}$ with $(w_h, w_{_{CP}})=(0, 20~{\rm GeV})$ are given in Fig.~\ref{fig:bubbleprofile}.
Our choice of the VEVs before the second step PT gives the maximal value of $\Gamma_{\rm sph}^{\rm br}$ where $\phi/T\simeq0.2$ as shown in Fig.~\ref{fig:GammaSphleron}, while $\Gamma_{\rm sph}^{\rm br}$ with $\kappa=0.1$ exceeds $\Gamma_{\rm sph}^{\rm sym}$ in that case.

The left panel of Fig.~\ref{fig:integrand} represents the sphaleron rate in the CKT prescription $\Gamma_\mathrm{\rm sph}^{\rm sym} f_\mathrm{sph}/T$ (orange) and $\Gamma_{\rm sph}^{\rm br}/T$ (green) for the benchmark scenario in Eq.~(\ref{eq:BMPT}).
Just for comparison we also show the result with the non-suppression sphaleron rate $\Gamma_{\rm sph}^{\rm sym}/T$ (blue). 
We can see the rate in the CKT prescription and $\Gamma_{\rm sph}^{\rm br}$ are almost the same in $z\lesssim0.1$~GeV$^{-1}$. However, the latter with $\kappa=10^{-1}$ (green dash-dot) is larger about one order in $z\gtrsim0.1$ GeV$^{-1}$ where $\phi /T\simeq0.2$. In contrast to $\kappa=10^{-1}$, the sphaleron rate with $\kappa=10^{-2}$ (green dotted) is almost the same as the CKT prescription case.
The right panel of Fig.~\ref{fig:integrand} indicates the integrand $I$ determined by $\eta_B=\int_0^\infty dz~I(z)$ as the function of the coordinate $z$ for the benchmark scenario.
The way to represent the results is the same as in the left panel.
We can see the integrands for the CKT and $\Gamma_{\rm sph}^{\rm br}$ are suppressed, and match in $z\lesssim0.1$ GeV$^{-1}$.
In $z\gtrsim0.1$ GeV$^{-1}$, the integrand with $\Gamma_{\rm sph}^{\rm br}$ with $\kappa=10^{-1}$ enhances because of the sphaleron increase shown in the left panel of Fig.~\ref{fig:integrand}. By contrast, the integrand with $\kappa=10^{-2}$ is nearly identical to the CKT case.
By integrating $I$ in Fig.~\ref{fig:integrand}, we can obtain the baryon asymmetries for the benchmark scenario as
\begin{align}
\eta_B \simeq 1.9\times10^{-10}
\,,\quad
\eta_{B}^{\mathrm{br}}(\kappa=10^{-1}) \simeq 1.8\times10^{-9}
\,,\quad
\eta_{B}^{\mathrm{br}}(\kappa=10^{-2}) \simeq 1.9\times10^{-10}
\,.
\label{eq:etaB}
\end{align}
In all cases, baryon asymmetries are produced more than the observed value.

To see the dependence of the phase position before the second step PT, we show the produced baryon number asymmetry in the $w_h$ vs. $w_{_{CP}}$ plane by using the CKT prescription in Fig.~\ref{fig:baryonasymmetery} and $\Gamma_{\rm sph}^{\rm br}$ in Fig.~\ref{fig:baryonasymmetery2}.\footnote{
We confirmed that the baryon asymmetries obtained by the FH scheme behave similarly to that from the CK scheme. The differences between the schemes are consistent with Ref.~\cite{Cline:2020jre}.
}
The colors in the plane indicate the produced baryon asymmetries when the second step PT occurs from the point $(w_h, w_{_{CP}})$ to the EW vacuum. The dashed lines represent the observed asymmetry, $\eta_B^{\rm obs}\simeq8.72\times 10^{-11}$. Typically, the region where $\phi=\sqrt{w_h^2+w_{_{CP}}^2} \simeq 20$--50 GeV in Figs.~\ref{fig:baryonasymmetery} and \ref{fig:baryonasymmetery2} can be consistent with the observation. 
In both figures, the baryon asymmetries gradually decrease in the upper right sides.
It is the consequence of the exponential suppression, which becomes effective for large $\phi /T$.
On the other hand, the asymmetries decrease on the bottom side of the figures. This is because the changes of the CP phase during the second step PT are small in the regions.
In Fig.~\ref{fig:baryonasymmetery2}, the asymmetries take the maximum value around $(w_h,~ w_{_{CP}})=(0,~ 20~{\rm GeV})$, which corresponds to our benchmark scenario.
In the case with $\kappa=10^{-2}$, where $\Gamma_{\rm sph}^{\rm br}$ hardly exceeds $\Gamma_{\rm sph}^{\rm sym}$ (cf.~Fig.~\ref{fig:GammaSphleron}), adequate baryon asymmetries can still be generated as in the right panel of Fig.~\ref{fig:baryonasymmetery2}.
Because $\kappa$ is an overall factor of the baryon asymmetry, the asymmetries between the panels in Fig.~\ref{fig:baryonasymmetery2} are different by one order.
Taking $\kappa\lesssim10^{-3}$, we find that the observed asymmetry cannot be reproduced in our scenario.
Note that if we change the value of $v_w$ to one, we get similar behaviors to the result in Ref.~\cite{Cline:2020jre}.
In conclusion, the PTs between the broken phases in the multi-step PTs can produce adequate baryon asymmetry in certain cases.

\section{Discussions and conclusion}\label{sec:5}
We can use the above results in multi-Higgs models with the $SU(2)_L$ charged scalar fields and the new CPV sources. 
For example, the CPV 2HDM, which has an extra $SU(2)_L$ scalar doublet, 
is a typical model for our scenario.
In the model with a softly broken $Z_2$ symmetry, the tree-level potential can be written as
\begin{align} 
\begin{split}\label{eq:tree_potential}
V_0 &= m_{1}^{2}\Phi_{1}^{\dagger}\Phi_{1} + m_{2}^{2}\Phi_{2}^{\dagger}\Phi_{2} - \left[m_{3}^{2}\Phi_{1}^{\dagger}\Phi_{2} + h.c.\right]
+ \frac{1}{2}\lambda_{1}(\Phi_{1}^{\dagger}\Phi_{1})^{2} + \frac{1}{2}\lambda_{2}(\Phi_{2}^{\dagger}\Phi_{2})^{2} \\ \relax
&\quad+ \lambda_{3}(\Phi_{1}^{\dagger}\Phi_{1})(\Phi_{2}^{\dagger}\Phi_{2}) + \lambda_{4}(\Phi_{1}^{\dagger}\Phi_{2})(\Phi_{2}^{\dagger}\Phi_{1})
+\left[\frac{1}{2}\lambda_{5}(\Phi_{1}^{\dagger}\Phi_{2})^{2} + h.c.\right]\,,
\end{split}
\end{align}
with scalar doublets,
\begin{align}\label{eq:scalar_doublet}
\Phi_{1} = \frac{1}{\sqrt{2}}
\begin{pmatrix}
\rho_{1}+i\eta_{1} \\
\phi_{1}+\zeta_{1}+i\psi_{1}
\end{pmatrix},\quad 
\Phi_{2} = \frac{1}{\sqrt{2}}
\begin{pmatrix}
\rho_{2}+i\eta_{2} \\
\phi_{2}+i\phi_{_{CP}}+\zeta_{2}+i\psi_{2}
\end{pmatrix}\,.
\end{align}
The doublets contain two neutral CP-even scalar fields $\phi_{1}$ and $\phi_{2}$, and one CP-odd field $\phi_{_{CP}}$, which can have VEVs.
Only considering these fields, the tree-level potential in Eq.~\eqref{eq:tree_potential} becomes
\begin{align}
\begin{split}
V_{\rm tree}(\phi_1,\phi_2,\phi_{_{CP}}) &= \frac{1}{2} m_{1}^2 \phi_{1}^2 + \frac{1}{2} m_{2}^2\left(\phi_{2}^2 + \phi_{_{CP}}^2\right) - \mathrm{Re}(m_{3}^2)\phi_{1}\phi_{2} + \mathrm{Im}(m_{3}^2)\phi_{1}\phi_{_{CP}}\\
&\quad + \frac{1}{8}\lambda_1\phi_{1}^4 + \frac{1}{8}\lambda_2\left(\phi_{2}^2 + \phi_{_{CP}}^2 \right)^2 +   \frac{1}{4}\lambda_3\phi_{1}^2\left(\phi_{2}^2 + \phi_{_{CP}}^2\right) + \frac{1}{4}\lambda_4\phi_{1}^2\left(\phi_{2}^2 + \phi_{_{CP}}^2\right) \\
&\quad+ \frac{1}{4}\mathrm{Re}(\lambda_5)\phi_{1}^2(\phi_{2}^2 - \phi_{_{CP}}^2) - \frac{1}{2}\mathrm{Im} (\lambda_5)\phi_{1}^2\phi_2\phi_{_{CP}}\,.
\end{split}
\end{align}
The CPV terms in the tree-level potential are 
$\mathrm{Im}(m_3^2)\phi_1\phi_{_{CP}}$ and $-\mathrm{Im}(\lambda_5)\phi_1^2\phi_2\phi_{_{CP}}/2$.
These two potential parameters relate to each other by the minimum condition as
$\mathrm{Im}(m_{3}^2) = v_1v_2\mathrm{Im}(\lambda_5)/2$, where $v_1$ and $v_2$ are the VEVs of $\phi_1$ and $\phi_2$ at the EW vacuum, respectively, with $\sqrt{v_1^2+v_2^2}=246$ GeV.
Hence, if $\mathrm{Im} (\lambda_5)>0$, the latter term 
gives negative contributions to the scalar potential and can make the CPV local minimum with non-zero $\phi_{_{CP}}$ 
at the finite temperature. On the other hand, when $\mathrm{Im} (\lambda_5)<0$, the former term 
can create the CPV minimum. One of the terms would enable the appropriate PT for our scenario.
In the following, we show the correspondence between the CPV 2HDM and our scenario if the suitable PT is realized.
Regarding the top Yukawa coupling, to prohibit the tree-level flavor changing neutral current, the top quark only couples $\Phi_2$.
Hence, the complex top quark mass becomes~\cite{Fromme:2006cm, Basler:2021kgq}
\begin{align}
m_t = \frac{1}{\sqrt{2}}Y_t(\phi_2+i\phi_{_{CP}})\equiv
\frac{1}{\sqrt{2}}Y_t\sqrt{\phi_2^2+\phi_{_{CP}}^2}e^{i\theta_{_{CP}}}\,,
\end{align}
with the top Yukawa coupling $Y_t$ in the 2HDM. 
The coupling $Y_t$ can be written by $y_t$ in Eq.~(\ref{eq:complexmt}) as $Y_t=y_t/\sin\beta$ with a mixing angle $\beta$ for the two scalar doublets.
The field $\phi_h$ and phase $\theta_{_{CP}}$ shown in Fig.~\ref{fig:bubbleprofile} can be rewritten as
\begin{align}
\phi_h\rightarrow\sqrt{\phi_1^2+\phi_2^2}\,, \quad \theta_{_{CP}}\rightarrow{\rm arg}\left(\phi_2+ i\phi_{_{CP}}\right)\,.
\end{align}
If we take $\tan\beta=1$ and replace parameters as above, we find that the sufficient baryon asymmetries can be generated in a second step PT which corresponds to our benchmark scenario in Eq.~(\ref{eq:BMPT}).
Therefore the sufficient baryon asymmetries can be generated in the multi-step PTs in the CPV 2HDM and its extensions. 
To verify the EWBG via the multi-step PTs phenomenologically, we need to find the specific PTs and consider constraints like the electron electric dipole moment (EDM) constraint~\cite{ACME:2018yjb}. 
As previous studies,
Ref.~\cite{Wang:2019pet}
found benchmark points for multi-step PTs in the CPV 2HDM, 
while the PTs would be difficult to make the observed BAU since the changes of $\phi_{_{CP}}$ are small.
Besides, for the one-step scenario, 
generated BAU in 2HDMs have been evaluated in
Refs.~\cite{Turok:1991uc, Cline:1995dg, Fromme:2006cm, Tulin:2011wi, Cline:2011mm, Liu:2011jh, Shu:2013uua, Chiang:2016vgf, Guo:2016ixx, Dorsch:2016nrg, Fuyuto:2017ewj, Modak:2018csw, Basler:2021kgq, 
Enomoto:2021dkl, Enomoto:2022rrl, Kanemura:2023juv}.
Ref.~\cite{Basler:2021kgq} suggests the difficulty of producing the adequate amount of the BAU 
using the WKB method.\footnote{Ref.~\cite{Basler:2021kgq} represents that the adequate BAU can be achieved using the VEV-insertion approximation. 
However, Ref.~\cite{Postma:2022dbr} shows the inaccuracy of the method.}
To obtain enough BAU in one-step PT in 2HDMs, a cancellation mechanism for suppressing
the electron EDM is necessary~\cite{Kanemura:2020ibp, Bian:2014zka, Cheung:2020ugr}. 
One example of such a mechanism
is proposed in Ref.~\cite{Kanemura:2020ibp}
for the general 2HDM, where new CPV phases are induced in the Higgs potential and Yukawa interaction, and
the electron EDM can be suppressed by the destructive interference between 
the fermion and the Higgs boson loop contributions in the Barr-Zee diagrams~\cite{Barr:1990vd}.
Along this line,
Refs.~\cite{Enomoto:2021dkl, Enomoto:2022rrl, Kanemura:2023juv}
have shown that there are some allowed regions for the successful EWBG in 2HDMs.\footnote{
Using a similar cancellation mechanism,
Ref.~\cite{Aoki:2022bkg} has shown a benchmark scenario for enough BAU in a radiative seesaw model,
that is an extension of the 2HDM.
}
Regarding the CP-conserving 2HDM, Ref.~\cite{Aoki:2021oez} shows that the multi-step PT favors the $Z_2$ symmetry in the tree-level potential
since the large $Z_2$ soft breaking parameter $\mathrm{Re}(m_3^2)$ leads to the one-step PT.
This would be the same as in the CPV 2HDM.
For the other studies on the PTs in the 2HDMs,
see, e.g., Refs.~\cite{Blinov:2015sna, Wang:2019pet, Fabian:2020hny, Aoki:2021oez, Benincasa:2022elt} for the multi-step PTs and, e.g., Refs.~\cite{Funakubo:1993jg, Cline:1996mga, Borah:2012pu, Gil:2012ya, Dorsch:2013wja, Ahriche:2015mea, Fuyuto:2015jha, Basler:2016obg, Dorsch:2017nza, Basler:2017uxn, Andersen:2017ika, Senaha:2018xek, Gould:2019qek, Su:2020pjw,Atkinson:2021eox, Goncalves:2021egx} for the one-step PTs.

The scenario, where the second step PT realizes the EWBG, was suggested in Ref.~\cite{Land:1992sm}, and its difficulty was represented in Ref.~\cite{Hammerschmitt:1994fn}.
They used a simplified 2HDM where the mixing term for the scalar doublets in the scalar potential is only $\lambda_3|\Phi^2||\eta|^2$ and the high-temperature potential, which includes thermal corrections only to the scalar quadratic terms.
Here, the $\Phi$ ($\eta$) is the SM-like (extra) scalar doublet that has the neutral CP-even scalar field $\phi_1$ ($\phi_2$).
Refs.~\cite{Land:1992sm, Hammerschmitt:1994fn} consider the two-step PT where the second step occurs from the point on the $\phi_2$ axis to that on the $\phi_1$ axis. It is a similar path of PT to our benchmark scenario in Eq.~(\ref{eq:BMPT}) by replacing $(\phi_h, \phi_{_{CP}})$ to $(\phi_1, \phi_2)$.
In the simple model, Ref.~\cite{Hammerschmitt:1994fn} found that the sphaleron is already suppressed before the second step PT occurs and concluded that sufficient baryon asymmetry via the two-step PT would be difficult. 
To verify the argument, we study the two-step PT in a more general model, the inert doublet model, using the effective potential in Refs.~\cite{Fabian:2020hny, Shibuya:2022xkj}.
The potential contains all the mixing terms allowed by the $Z_2$ symmetry ($\Phi\rightarrow\Phi$, $\eta\rightarrow-\eta$) and the renormalizability, while some of them are taken to be zero in Ref.~\cite{Hammerschmitt:1994fn}.
The potential also includes contributions from the one-loop and the daisy diagrams. In contrast to Ref.~\cite{Hammerschmitt:1994fn}, using CosmoTransitions~\cite{Wainwright:2011kj}, we find the PTs where the sphaleron rates before the second step PT would be hardly suppressed.
For example, we find the benchmarks (BMs)\footnote{
The BMs satisfy the constraints from the perturbativity, the bound from below, the unitarity, the electroweak precision data, and collider data from LEP and LHC~\cite{Pierce:2007ut, Cao:2007rm, Lundstrom:2008ai, Krawczyk:2013pea, Abouabid:2020eik}.
In terms of DM, it is difficult to satisfy the direct detection constraints~\cite{XENON:2018voc,PandaX-4T:2021bab, LZ:2022ufs}
if the lightest inert particle (LIP) is the only DM candidate.
However, there are possibilities to avoid the constraints
by extending the model, such as
considering multi-component DM scenarios
where the LIP is part of a multi-component DM~\cite{Alves:2016bib, Aoki:2017eqn, Chakraborti:2018aae, Bhattacharya:2019fgs} or introducing a new $Z_2$-odd particle (e.g. $Z_2$-odd right-handed neutrino
in Ref.~\cite{Ma:2006km}) lighter than the LIP as a DM.}
where the second step PTs occur in the following paths,
\begin{align*}
&\mathrm{BM1:}\quad
(\phi_1, \phi_2)=(0,~13.0~{\rm GeV})~\rightarrow~(240~{\rm GeV},~0)~~\mathrm{~~at}~~T_n=\mathrm{46.8~GeV}\,,\\
&\mathrm{BM2:}\quad
(\phi_1, \phi_2)=(0,~19.1~{\rm GeV})~\rightarrow~(222~{\rm GeV},~0)~~\mathrm{~~at}~~T_n=\mathrm{67.2~GeV}\,,
\end{align*}
with the bubble nucleate temperature $T_n$ at the parameter sets,
\begin{align*}
&\mathrm{BM1:}\quad
m_{\eta^\pm}=m_A\simeq437~\mathrm{GeV},
\quad m_H\simeq65.7~\mathrm{GeV}, \quad
\lambda_2 \simeq2.00, \quad \mu_\eta^2\simeq-5.31\times10^{3}~\mathrm{GeV}^2\,,\\
&\mathrm{BM2:}\quad
m_{\eta^\pm}=m_A\simeq398~\mathrm{GeV},
\quad m_H\simeq513~\mathrm{GeV}, \quad
\lambda_2 \simeq3.31, \quad \mu_\eta^2\simeq2.19\times10^{4}~\mathrm{GeV}^2\,.
\end{align*}
The masses $m_{\eta^\pm}, m_A$, and $m_H$ correspond to the charged, neutral CP-odd, and CP-even inert scalar fields, respectively. The indexes $\lambda_2$ and $\mu_\eta^2$ are scalar potential parameters, and those potential terms are $\mu_\eta^2\phi_2^2$ and $\lambda_2\phi_2^4$. The other parameters can be calculated using equations in Ref.~\cite{Shibuya:2022xkj}.
The value of $\phi/T~(=\sqrt{\phi_1^2+\phi_2^2}/T)$ before the second step for each BM is 0.278 (BM1) and 0.285 (BM2). We can see that sphaleron rates in both BMs would be hardly suppressed as shown in Fig.~\ref{fig:GammaSphleron}. Hence, suitable PTs whose sphaleron rates are not suppressed before the second step PT can exist contrary to Ref.~\cite{Hammerschmitt:1994fn}.
Note that in both BMs, the wall widths $L_w$ are about 0.1 GeV$^{-1}$, which is the same as in our benchmark scenario in Sec.~\ref{sec:4}.

In the inert doublet model, CPV terms are prohibited since the scalar potential has the exact $Z_2$ symmetry. 
Hence, higher dimensional operators have been considered to make the new CPV sources as 
$(C_1\Phi^\dagger\Phi + C_2\eta^\dagger\eta)\mathrm{tr}(W_{\mu\nu}\widetilde{W}^{\mu\nu})/\Lambda^2$~\cite{Dine:1990fj, Gil:2012ya, Fabian:2020hny},
where the index $\Lambda$ is an UV cut-off scale, and $W_{\mu\nu}$ and $\widetilde{W}^{\mu\nu}$ are the $SU(2)_L$ field strength and the dual, respectively.
The coefficients $C_1$ and $C_2$ determine the largeness of the CPV.
The $C_1$ is constrained strictly by the electron EDM~\cite{Fabian:2020hny}. Therefore, a case with small enough $C_1$ and large enough $C_2$ for realizing the observed BAU would be hopeful.

In this letter, we calculated the produced baryon number asymmetry via the second step PT in the multi-step EWPT using the prototypical model. Our results and discussions can be widely applied in models with $SU(2)_L$ extra scalar fields, which can have the VEVs, and new CPV sources. 
Since we consider the PT occurring from the $SU(2)_L$ broken phase to the broken phase, we utilize the sphaleron rates in the prescription suggested in Ref.~\cite{Cline:2011mm} and the broken phase~\cite{Arnold:1987mh}.
As a result, we show that adequate baryon number asymmetries can be produced in both cases.
In particular, the observed asymmetry can be reproduced for $\sqrt{w_h^2+w_{_{CP}}^2}\simeq20$--50 GeV, where $w_h$ and $w_{_{CP}}$ are the scalar VEVs before the second step PT.
In addition, we discuss the EWBG in the CPV 2HDM and find that the sufficient asymmetries are possible to be realized, although phenomenological constraints are not considered.
Meanwhile, in the inert doublet model, we discover suitable multi-step PTs where the sphaleron rates would not be suppressed before the second step PTs in contrast to the previous study in Ref.~\cite{Hammerschmitt:1994fn}.
We expect that the sphaleron rate before the second step PT between broken phases can be hardly suppressed in the model with extra $SU(2)$ scalar fields.
In addition, if there are enough CPV sources, the adequate baryon number asymmetry would be produced.
We would like to mention that we need non-perturbative calculations to obtain a precise sphaleron rate in the region with small $\phi/T$.
Progresses in the non-perturbative studies are essential for correctly evaluating the baryon number asymmetry via the EWBG between the broken phases in the multi-step PT.

\appendix
\section{Transport equations}\label{sec:A}
The transport equations for the chemical potentials of the top quarks $\mu_t$ and $\mu_{t^c}$, left-handed bottom quarks $\mu_b$, 
Higgs doublets $\mu_h$, and the relevant plasma velocities $u_k$ ($k=t,b,t^c,h$) in the CK scheme are~\cite{Cline:2020jre, Enomoto:2022rrl}
\begin{equation}
\begin{aligned}
-S_{1,t}=&~
-D_{1,t} \mu_t^\prime + u_t^\prime + \gamma_w v_w (m_t^2)^\prime Q_{1,t} \mu_t - K_{0,t} \overline{\Gamma}_t 
\,\,,\\[.1cm]
0=&~
-D_{1,b} \mu_b^\prime + u_b^\prime - K_{0,b} \overline{\Gamma}_b 
\,\,,\\[.1cm]
-S_{1,t}=&~
-D_{1,t} \mu_{t^c}^\prime + u_{t^c}^\prime + \gamma_w v_w (m_t^2)^\prime Q_{1,t} \mu_{t^c} - K_{0,t} \overline{\Gamma}_{t^c}
\,\,,\\[.1cm]
0=&~
-D_{1,h} \mu_h^\prime + u_h^\prime - K_{0,h} \overline{\Gamma}_h 
\,\,,\\[.1cm]
-S_{2,t}=&~
-D_{2,t} \mu_{t}^\prime - v_w u_{t}^\prime + \gamma_w v_w (m_t^2)^\prime Q_{2,t} \mu_t + (m_t^2)^\prime \overline{R}_t u_{t} + \Gamma_{t,\mathrm{tot}} u_t + v_w K_{0t} \overline{\Gamma}_t 
\,\,,\\[.1cm]
0=&~
-D_{2,b} \mu_b^\prime - v_w u_b^\prime + \Gamma_{b,\mathrm{tot}} u_b + v_w K_{0,b} \overline{\Gamma}_b 
\,\,,\\[.1cm]
-S_{2,t}=&~
-D_{2,t} \mu_{t^c}^\prime - v_w u_{t^c}^\prime + \gamma_w v_w (m_t^2)^\prime Q_{2,t} \mu_{t^c} + (m_t^2)^\prime \overline{R}_t u_{t^c} + \Gamma_{t,\mathrm{tot}} u_{t^c} + v_w K_{0,t} \overline{\Gamma}_{t^c}
\,\,,\\[.1cm]
0=&~
-D_{2,h} \mu_h^\prime - v_w u_h^\prime + \Gamma_{h,\mathrm{tot}} u_h + v_w K_{0,h} \overline{\Gamma}_h \,,
\end{aligned}
\end{equation}
with the source terms for the top quark,
\begin{equation}
S_{l,t} = -\gamma_w v_w (m_t^2 \theta_{_{CP}}^\prime)^\prime Q_{l,t}^8 + \gamma_w v_w m_t^2 \theta_{_{CP}}^\prime(m_t^2)^\prime Q_{l,t}^9 \quad (l=1,2)\,.
\end{equation}
The prime means the derivative by the coordinate $z$.
The functions $D$, $Q$, $K_0$, and $\overline{R}$ are given in Ref.~\cite{Cline:2020jre}. 
There is a typo for $K_0$ in Ref.~\cite{Cline:2020jre} as Ref.~\cite{Lewicki:2021pgr} indicates.
Hence, we include a factor $1/T$ in $K_0$. 
The inelastic reaction rate for each particle are~\cite{Cline:2020jre}
\begin{equation}
\begin{aligned}
\overline{\Gamma}_t &= \Gamma_{_{SS}} \Bigl((1 + 9D_{0,t}) \mu_t + 10 \mu_b + (1-9D_{0,t}) \mu_{t^c} \Bigr) \\
&~~+ \Gamma_{_W} (\mu_t - \mu_b) + \Gamma_y (\mu_t + \mu_{t^c} + \mu_h) +2 \Gamma_m (\mu_t + \mu_{t^c})\,,\\
\overline{\Gamma}_b &= \Gamma_{_{SS}} \Bigl( (1 + 9D_{0,t}) \mu_t + 10 \mu_b + (1 + 9D_{0,t}) \mu_{t^c} \Bigr) \\
&~~+\Gamma_{_W} (\mu_b - \mu_t) + \Gamma_y (\mu_b + \mu_{t^c} + \mu_h)\,,\\
\overline{\Gamma}_{t^c} &= \Gamma_{_{SS}} \Bigl( (1 + 9D_{0,t}) \mu_t + 10 \mu_b + (1 - 9D_{0,t}) \mu_{t^c} \Bigr) \\
&~~+ 2 \Gamma_m ( \mu_{t^c} + \mu_t) + \Gamma_y (2 \mu_{t^c} + \mu_t + \mu_b + 2 \mu_h)\,,\\
\overline{\Gamma}_h &= \frac{3}{4} \Gamma_y (2 \mu_h + \mu_t + \mu_b + 2\mu_{t^c}) + \Gamma_h \mu_h,
\end{aligned}
\end{equation}
with
\begin{align}
\Gamma_{_{SS}}=4.9\times10^{-4}T\,,\quad
\Gamma_y=4.2\times10^{-3}T\,,\quad
\Gamma_m=m_t^2(z)/(63T)\,,\quad
\Gamma_h=m_{_W}^2(z)/(50T),
\end{align}
and the W boson mass given by $m_{_W}=g\phi /2$.
The definition of $\Gamma_W$ is in Ref.~\cite{Cline:2020jre}.
Following Ref.~\cite{Cline:2020jre}, we take the masses of the bottom quarks and Higgs as massless because their impacts are small~\cite{Fromme:2006wx, Fromme:2006cm}.
Using the chemical potentials derived by the transport equations, the chemical potential for the left-handed baryons can be calculated as
\begin{align}\label{eq:muBL}
\mu_{B_L}
&=\frac{1}{2}(1+4D_{0,t})\mu_{t}+\frac{1}{2}(1+4D_{0,b})\mu_{b}
-2D_{0,t}\mu_{t^c}\,.
\end{align}

\begin{acknowledgments}
HS would like to thank Chikako Idegawa for the valuable discussion and Kazuki Enotomo and Mura Yushi for their helpful correspondence. 
The work of M.~A. is supported in part by the Japan Society for the Promotion of Sciences Grant-in-Aid for Scientific Research (Grant No. 20H00160). 
The work of H.~S. is supported by JST SPRING, Grand No.~JPMJSP2135.
\end{acknowledgments}

\let\doi\relax
\bibliographystyle{JHEP}
\bibliography{Ref}

\end{document}